\documentclass[11pt]{article}
\usepackage{amsmath,amssymb,color,graphics,epsfig}

\usepackage{amsfonts}

\newcommand{\be}{\begin{equation}}
\newcommand{\ee}{\end{equation}}
\newcommand{\bea}{\setlength\arraycolsep{2pt} \begin{eqnarray}}
\newcommand{\eea}{\end{eqnarray}}
\newcommand{\nn}{\nonumber}

\def\ft#1#2{{\textstyle{\frac{\scriptstyle #1}{\scriptstyle #2} } }}
\def\fft#1#2{{\frac{#1}{#2}}}

\def\0{{\sst{(0)}}}
\def\1{{\sst{(1)}}}
\def\2{{\sst{(2)}}}
\def\3{{\sst{(3)}}}
\def\4{{\sst{(4)}}}
\def\5{{\sst{(5)}}}
\def\6{{\sst{(6)}}}
\def\7{{\sst{(7)}}}
\def\8{{\sst{(8)}}}
\def\sst#1{{\scriptscriptstyle #1}}

\begin{document}

\begin{flushright}
\end{flushright}

\vspace{25pt}
\begin{center}
{\large {\bf Critical phenomena of regular black holes in anti-de Sitter space-time}}

\vspace{10pt}
 Zhong-Ying Fan

\vspace{10pt}
{\it Center for High Energy Physics, \\}
{\it Peking University, No.5 Yiheyuan Rd, Beijing 100871, P. R. China\\}

\vspace{40pt}

\underline{ABSTRACT}
\end{center}
In General Relativity coupled to a non-linear electromagnetic field, together with a negative cosmological constant, we obtain the general static spherical symmetric black hole solution with magnetic charges, which is asymptotic to anti-de Sitter (AdS) space-times. In particular, for a degenerate case the solution becomes a Hayward-AdS black hole, which is regular everywhere in the full space-time. The existence of such a regular black hole solution preserves the weak energy condition while the strong energy condition is violated. We then derive the first law and the Smarr formula of the black hole solution. We further discuss its thermodynamic properties and study the critical phenomena in the extended phase space where the cosmological constant is treated as a thermodynamic variable as well as the parameter associated with the non-linear electrodynamics. We obtain many interesting results such as: the Maxwell's equal area law in the $P-V$ (or $S-T$) diagram is violated and consequently the critical point $(T_*\,,P_*)$ of the first order small-large black hole transition does not coincide with the inflection point ($T_c\,,P_c$) of the isotherms; the Clapeyron equation describing the coexistence curve of the Van der Waals (vdW) fluid is no longer valid; the heat capacity at constant pressure is finite at the critical point; the various exponents near the critical point are also different from those of the vdW fluid.

\vfill {\footnotesize  Email: fanzhy@pku.edu.cn\,.}

\thispagestyle{empty}

\pagebreak

\tableofcontents
\addtocontents{toc}{\protect\setcounter{tocdepth}{2}}



\section{Introduction}
Black hole is one of the most important objects predicted by General Relativity. A mysterious property of black hole is that all the first exact black hole solutions known in General Relativity have a singularity at the origin of the space-time. In fact, the celebrated singularity theorems proved by Penrose and Hawking \cite{hawking} state that under some physically reasonable conditions the existence of singularities is inevitable in General Relativity. However, it is widely believed that the space-time singularities reflect the limitation of classical theories of gravity and can be avoided in nature when quantum effects are considered. This refers to a definite theory of quantum gravity. As such a theory has not yet been well developed, it is instructive to consider how to avoid black hole singularity at the semi-classical level.

The first regular black hole model (mostly known as ``Bardeen black hole") is proposed by Bardeen \cite{bardeen}.
A straightforward analysis shows that the Bardeen black hole is indeed free of singularity: the singularity at the origin of the space-time is replaced by a de Sitter patch. The price of having such a regular model is that some physical conditions for ordinary matter fields such as the strong energy condition are violated\footnote{For rotating regular black holes, the weak energy condition will also be violated as far as we are aware. }. Other regular black hole models were also proposed in the literature   \cite{Borde:1994ai,Barrabes:1995nk,Cabo:1997rm,Hayward:2005gi,Bambi:2013ufa,Ghosh:2014hea,Toshmatov:2014nya,Azreg-Ainou:2014pra}. It was much later realized by Ay\'{o}n-Beato and Garc\'{i}a   \cite{AyonBeato:1998ub,AyonBeato:1999ec,AyonBeato:1999rg,AyonBeato:2000zs} that the physical source of regular black holes could be a nonlinear electrodynamics. In particular, it was shown in \cite{AyonBeato:2000zs} that the Bardeen black hole can be interpreted as the gravitational field of a nonlinear magnetic monopole\footnote{The Bardeen black hole in the gravity model considered in \cite{AyonBeato:2000zs} contains one free integration constant. This is not apparent due to the parametrization of that paper.}.  Recently, some regular black hole solutions have also been constructed in $f(T)$ gravity coupled to a nonlinear electrodynamics \cite{Junior:2015fya}.

In this paper, motivated by the pioneer work of Ay\'{o}n-Beato and Garc\'{i}a, we would like to further study Einstein gravity coupled to a nonlinear electromagnetic field. For later purpose, we also introduce a negative cosmological constant. We successfully construct a well known regular black hole model, the Hayward black hole \cite{Hayward:2005gi} and its generalization in AdS space-times for a certain nonlinear electromagnetic field. The solution carries magnetic charges and contains one free integration constant. Thus, the physical interpretation of the Hayward black hole follows the Bardeen black hole constructed in \cite{AyonBeato:2000zs}: it is a degenerate configuration of the gravitational field of a non-linear magnetic monopole. The general static spherical symmetric black hole solution involves an extra {\it Schwarzschild mass} term such that the solution reduces to a Schwarzschild black hole in the neutral limit.

We study the global properties of the solution and derive the first law of thermodynamics. Treating the cosmological constant as a thermodynamic variable as well as the parameter associated with the nonlinear electromagnetic field, we also derive the generalized first law in the extended phase space. We then study the critical phenomena of the regular black hole in the extended phase space and obtain many intriguing results which are different from those of the Van der Waals (vdW) fluid.

The paper is organized as follows. In section 2, we study Einstein gravity coupled to a nonlinear electromagnetic field and obtain Hayward-AdS black hole with magnetic charges. We study the global properties of the solution and derive the corresponding first law. In section 3, by taking the parameter $\sigma$ associated with the nonlinear electromagnetic field as a dynamic variable, we study the $P-V$ criticality of the black hole solution in the extended phase space. In section 4, we briefly discuss the critical phenomena of the general black hole solution with fixed $\sigma$ and magnetic charge. We conclude this paper in section 5.

\section{Einstein gravity coupled to non-linear electrodynamics}\label{sec2}
We consider Einstein gravity coupled to a non-linear electromagnetic field of the type
\be I=\fft{1}{16\pi}\int \mathrm{d}^4x\sqrt{-g}\, \Big(R+6\ell^{-2}-\mathcal{L}(\mathcal{F})\Big) \,,\ee
where $\mathcal{F}\equiv F_{\mu\nu}F^{\mu\nu}$, $F=dA$ is the field strength of the electromagnetic field and $\mathcal{L}$ is a function of $\mathcal{F}$. The covariant equations of motion are
\be
G_{\mu\nu}=T_{\mu\nu}\,,\qquad \nabla_{\mu}\Big( \mathcal{L}_{\mathcal{F}} F^{\mu\nu}\Big)=0 \,,
\label{eom}\ee
where $G_{\mu\nu}=R_{\mu\nu}-\fft 12 (R+6\ell^{-2}) g_{\mu\nu}$ is the Einstein tensor and $\mathcal{L}_{\mathcal{F}}=\fft{\partial\mathcal{L}}{\partial\mathcal{F}} $. The energy momentum tensor is
\be T_{\mu\nu}=2\Big( \mathcal{L}_{\mathcal{F}} F_{\mu\nu}^2-\fft 14 g_{\mu\nu} \mathcal{L} \Big)  \,.\label{energymomentum}\ee
In this paper, we consider static spherical symmetric black holes with magnetic charges. The most general ansatz is given by
\be ds^2=-f dt^2+\fft{dr^2}{f}+r^2 d\Omega^2 \,,\qquad A=Q_m \cos{\theta}\, d\phi \,,\label{ansatz}\ee
where $f=f(r)$ and $d\Omega=d\theta^2+\sin{\theta}^2 d\phi^2$ denotes the metric of a unit $2$-sphere, $Q_m$ is the total magnetic charge carried by the black hole
\be Q_m=\fft{1}{4\pi}\int_{\Sigma_2} F \,.\label{mcharge}\ee
In \cite{fangw}, a general strategy will be developed for constructing exact black hole solutions with electric/magnetic charges in this gravity model. Here for our purpose, we focus on a well known regular black hole model, namely the Hayward black hole \cite{Hayward:2005gi} generalized in AdS space-time.
\subsection{Hayward-AdS black hole}
\begin{figure}[ht]
\centering
\includegraphics[width=230pt]{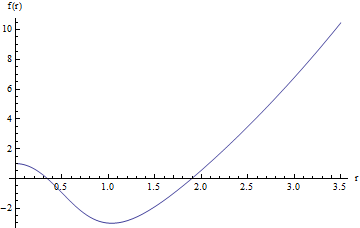}
\caption{{\it The plot of the metric function $f(r)$ for Hayward-AdS black hole with $\ell=1\,,q=1\,,\sigma=1/5$. In this case, the black hole has two horizons. For appropriate parameters, the black hole can also have only one horizon and become extremal.}}
\label{fig0}\end{figure}

It turns out that for the Lagrangian density
\be \mathcal{L}=12 \sigma^{-1}\fft{\big(\sigma \mathcal{F} \big)^{3/2}}
{\Big(1+\big(\sigma \mathcal{F} \big)^{3/4} \Big)^2} \,,\label{lagrangian1}\ee
we can obtain the Hayward black hole in AdS space-time
\bea\label{sol1}
&&ds^2=-f dt^2+\fft{dr^2}{f}+r^2 d\Omega^2 \,,\qquad A=Q_m \cos{\theta}\, d\phi \,,\nn\\
&&f=r^2/\ell^2+1-\fft{2\sigma^{-1} q^3 r^2}{ r^3+q^3}\,,
\eea
where $q$ is an integration constant which is related to the magnetic charge.
Note that in the weak field limit the vector field becomes $\mathcal{L}\sim \sigma^{1/2}\mathcal{F}^{3/2}$, which is stronger than a linear Maxwell field. In Fig. \ref{fig0}, we plot the metric function function $f(r)$. It is easy to see that for proper parameters there exist one or two horizons corresponding to the positive real roots of the equations $f(r)=0$ and $f$ approaches unity at the origin of the space-time.
In fact, near the origin the metric function behaves as
\be f=1+(\ell^{-2}-2\sigma^{-1})r^2+\cdots \,.\ee
Thus, depending on the parameters of the theory the metric is either asymptotically Minkowskian or (A)dS space-times at the origin. To ensure the geometry is
indeed regular at the origin, we calculate some low-lying curvature polynomials and find that all of them have a finite value at this point
\be R=-12\big(\ell^{-2}-2\sigma^{-1}\big)\,,\quad R^2_{\mu\nu}=36\big(\ell^{-2}-2\sigma^{-1}\big)^2\,,\quad  R^2_{\mu\nu\rho\sigma}=24\big(\ell^{-2}-2\sigma^{-1}\big)^2 \,.\ee
 In fact, the geometry is regular everywhere in the space-time. This is significantly different from the conventional black holes such as the Schwarzschild and Reissner-Nordstr\"{o}m black holes which in general have a singularity at the origin. The price that we pay for obtaining such a regular black hole is that the strong energy condition is violated. Nevertheless, the weak energy condition is still preserved.

It should be emphasized that in our theory the Hayward-AdS black hole (\ref{sol1}) contains only one independent integration constant and hence is a degenerate solution. The general two-parameter family black hole solution reads
\bea\label{sol2}
&&ds^2=-f dt^2+\fft{dr^2}{f}+r^2 d\Omega^2 \,,\qquad A=Q_m \cos{\theta}\, d\phi \,,\nn\\
&&f=r^2/\ell^2+1-\fft{2M}{r}-\fft{2\sigma^{-1} q^3 r^2}{ r^3+q^3}\,,
\eea
where $M$ is associated with the condensate of the massless graviton, originating from its self-interactions. In the neutral limit, the solution reduces to the Schwarzschild-AdS black hole. Hence, we refer to $M$ as {\it Schwarzschild mass}. It is clear that for any non-zero $M$, the metric behaves singular at the origin and the existence of the singularity is unavoidable.

\subsection{The first law of thermodynamics}
In this subsection, we will derive the first law of thermodynamics for the Hayward-AdS black hole. For the discussion to be as simple as possible, we focus on the general black hole solution (\ref{sol2}). At asymptotic infinity, the metric function behaves as
\be f=r^2/\ell^2+1-\ft{2\big(M+\sigma^{-1}q^3\big)}{r}+\cdots \,,\ee
from which we can read off the AMD mass  \cite{Ashtekar:1984zz,Ashtekar:1999jx}
\be M_{\mathrm{AMD}}=M+\sigma^{-1}q^3 \,.\label{adm}\ee
It is interesting to note that the AMD mass or equivalently the condensate of the massless graviton, has two copies of contributions, one from the self-interactions of the graviton, giving rise to the {\it Schwarzschild mass} and the other from the non-linear interactions between the graviton and the (non-linear) photon, leading to the charged term $\sigma^{-1}q^3$. The latter contribution is impossible for a linear Maxwell field.

The temperature and entropy are given by
\be T=\fft{3r_0^6+\Big( r_0(r_0^3-2q^3) +6M q^3\Big)\ell^2}{4\pi r_0^2(r_0^3+q^3)\ell^2}\,,\qquad S=\pi r_0^2 \,,\label{tem1}\ee
where $r_0$ is the horizon radius defined by the largest root of the equation $f(r_0)=0$.
The magnetic charge defined by (\ref{mcharge}) is
\be Q_m=\fft{q^2}{\sqrt{2\sigma}} \,.\label{magnetic}\ee
whilst the conjugate potential should be redefined properly. A generalized definition was provided in \cite{Rasheed:1997ns}
\be \Psi=\widetilde{A}_t(r_0)-\widetilde{A}_t(\infty)\,,\qquad \widetilde{F}=d\widetilde{A}=\mathcal{L}_{\mathcal{F}}\, {}^*F \,,\ee
which coincides with the conventional one for a linear Maxwell filed. We find
\be \Psi=\fft{3q^4(2r_0^3+q^3)}{\sqrt{2\sigma}\,(r_0^3+q^3)^2} \,.\ee
Then the standard first law
\be dM_{\mathrm{AMD}}=T dS+\Psi dQ_m \,,\ee
holds straightforwardly.  It is worth pointing out that in \cite{Rasheed:1997ns}, the first law of asymptotically flat black holes with nonlinear electric/magnetic charges was derived from a covariant approach. In the extended phase space where the cosmological constant and the parameter $\sigma$ of the non-linear electromagnetic filed are taken as thermodynamic variables, the first law is generalized to
\be dM_{\mathrm{AMD}}=T dS+\Psi dQ_m+V dP+\Pi d\sigma \,,\label{firstlaw}\ee
where the pressure and the thermodynamic volume are defined as usual
\cite{Kastor:2009wy,Cvetic:2010jb}
\be P=-\fft{\Lambda}{8\pi}=\fft{3}{8\pi\ell^2}\,,\qquad V=\fft{4\pi r_0^3}{3} \,,\ee
and a new quantity $\Pi$ conjugate to $\sigma$ is introduced. It is defined by
\be \Pi=\fft 14\int_{r_0}^{\infty}\mathrm{d}r\,\sqrt{-g}\,\fft{\partial\mathcal{L}}{\partial\sigma} \,.\label{Pi}\ee
We find
\be \Pi=\fft{q^6(2r_0^3-q^3)}{4\sigma^2(r_0^3+q^3)^2} \,.\ee
The Smarr formula turns out to be
\be M_{\mathrm{AMD}}=2TS+\Psi Q_m-2VP+2\Pi \sigma \,,\label{smarr}\ee
 which is perfectly consistent with the scaling dimensional argument. It should be emphasized that the existence of the new pair of conjugates $(\Pi\,,\sigma)$ is essential to govern the validity of the Smarr formula but the definition of the quantities $(\Pi\,,\sigma)$ is not unique. One can properly define a new quantity $\widetilde{\sigma}\propto\sigma^z$ and the conjugate variable as $\widetilde{\Pi}d\widetilde{\sigma}=\Pi d\sigma$. Then the Smarr formula (\ref{smarr}) is still valid with $2\Pi \sigma$ term replaced by $2z \widetilde{\Pi}\widetilde{\sigma}$. An interesting question is how to interpret the new pair of conjugates in physics. Unfortunately, unlike the Born-Infeld case the parameter $\sigma$ (or its some power law) does not have a preferred interpretation for our solution. Hence, the question remains open.

 Finally, to end this section, we remark that for the Hayward black hole that is asymptotic to Minkowskian space-times, there are three roots for the equation $f(r)=0$. Thus, mathematically the black hole has three horizons (two real and one imaginary or one real and two imaginary or three imaginary) and their product of entropies is
 \be \prod_{\alpha=1}^{3}S_\alpha=(2\sigma)^{3/2}\pi^3Q_m^3 \,,\ee
 which is intriguingly expressed in terms of the magnetic charge and the parameter $\sigma$. Moreover, this relation is independent of the specific details of the black hole mass as a function of $Q_m$ and $\sigma$. The universality of this property may provide some insights for probing the microscopics of regular black holes.

\section{P-V criticality of Hayward-AdS black hole}
\begin{figure}[ht]
\centering
\includegraphics[width=230pt]{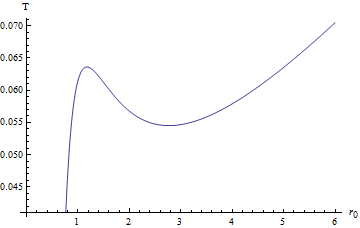}
\caption{{\it The plot of the temperature as a function of the horizon radius for Hayward-AdS black hole with $\ell=5\,,q=1/2$. For any given temperature between the local minimum and maximum, there are three black holes with different horizon radius. }}
\label{fig00}\end{figure}

For Hayward-AdS black hole (\ref{sol1}), the temperature is given by (\ref{tem1}) with zero {\it Schwarzschild mass}
\be T=\fft{3r_0^5+(r_0^3-2q^3)\ell^2}{4\pi r_0(r_0^3+q^3)\ell^{2}} \,.\label{tem2}\ee
Choosing the parameters appropriately, we find that there are various black holes with different horizon radius for a certain range of the temperature (see Fig. \ref{fig00}). This indicates that probably like the Reissner-Nordstr\"{o}m  black hole case \cite{Kubiznak:2012wp,Gunasekaran:2012dq} there exists a small-large black hole (SBH-LBH) transition in the extended phase space. We will show that this is indeed true. The critical phenomena of other AdS black holes with non-linear electromagnetic charges have also been discussed in the literature \cite{Gunasekaran:2012dq,Banerjee:2011cz,Banerjee:2012zm,Mo:2014qsa,Mo:2016jqd,Hendi:2014kha}.
\subsection{Equation of state and Gibbs free energy}
\begin{figure}[ht]
\includegraphics[width=210pt]{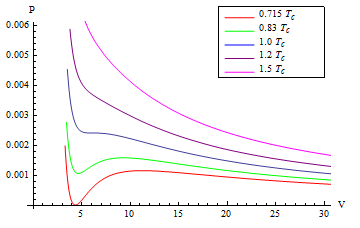}
\includegraphics[width=210pt]{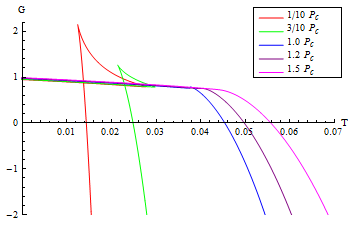}
\caption{{\it The left plot is the P-V diagram for various temperatures near the critical point. The right plot is the Gibbs free energy as a function of the temperature for different pressures. We have set $q=1$.}}
\label{fig1}\end{figure}

As mentioned earlier, the Hayward-AdS black hole (\ref{sol1}) in our case is a degenerate solution. In order to study its critical phenomena, we shall take the parameter $\sigma$ as a dynamical variable (This is said in the thermodynamic sense: $\sigma$ is changed during the phase transition.). Then the equation (\ref{tem2}) gives rise to the equation of state
\be P=\fft{T}{v}-\fft{1}{2\pi v^2}+\fft{8T q^3}{v^4}+\fft{8q^3}{\pi v^5} \,.\ee
where $v=2\ell_p^2\, r_0$ is the specific volume, $\ell_p$ is Plank length which will be set to unity throughout this paper. In addition, owing to the first law (\ref{firstlaw}), the AMD mass of the black hole should be identified with the enthalpy of the dual fluid. Thus, by definition, the Gibbs free energy is
 \be G=M_{\mathrm{AMD}}-T S \,.\ee
 Alternatively, we also calculate the Euclidean action (plus surface terms and counter terms) and obtain the same result. In Fig. \ref{fig1},
we plot the $P-V$ diagram for different temperatures in the left panel and the Gibbs free energy as a function of the temperature for various pressures in the right panel, respectively.
We see that below a critical temperature, there exists an oscillating part in the isotherms of the $P-V$ diagram. Moreover, the characteristic ``swallow-tail" behavior of the Gibbs free energy is a strong indication that below a critical temperature, a first order SBH-LBH transition can occur at the intersection point of the Gibbs free energy. This confirms our naive expectation at the beginning of this section. A critical point occurs when there exists an inflection point in the $P-V$ diagram, namely
\be \fft{\partial P}{\partial v}=0\,,\qquad \fft{\partial^2 P}{\partial v^2}=0 \,.\label{crit}\ee
This is the critical point at which one cannot clearly distinguish between SBH and LBH phases. For our case, we obtain
\bea\label{crit1}
P_c&=&\fft{3(57-23\sqrt{6})(14+6\sqrt{6})^{1/3}}{800\pi q^2}\,,\nn\\
T_c&=&\fft{(5-2\sqrt{6})(14+6\sqrt{6})^{2/3}}{8\pi q}\,,\nn\\
v_c&=&2(14+6\sqrt{6})^{1/3}\,q \,.\eea
Thus
\be \fft{P_c v_c}{T_c}=\fft{3(9-\sqrt{6})}{50}\approx 0.3930 \,,\ee
is a universal constant which is slightly bigger than the value $3/8$ of the vdW fluid. It should be emphasized that this critical point does not necessarily coincide with the turning point at which the first order transition terminates. The latter should be determined by the physical conditions associated with a first order transition. This will be discussed in details in the next subsection.

Finally, for later convenience, we define some dimensionless quantities
\be p=P/P_c\,,\qquad \tau=T/T_c\,,\qquad \nu=v/v_c \,.\ee
Then the equation of state leads to the so-called law of corresponding states
\be \ft 92 p \,\nu^2=\Big( (9+\sqrt{6})\nu+\ft{4\sqrt{6}-9}{2\nu^2} \Big)\tau-(3+2\sqrt{6})+\ft{3-\sqrt{6}}{\nu^3}  \,.\ee
In addition, the Gibbs free energy can be expressed as $G(T\,,P)=q\, \mathcal{G}(\tau\,,p)$, where $\mathcal{G}$ is a function of $\tau\,,p$ and also dimensionless. These relations involving only dimensionless quantities will be very useful in later calculations.

\subsection{Area law, critical point and phase diagram}
\begin{figure}[ht]
\includegraphics[width=210pt]{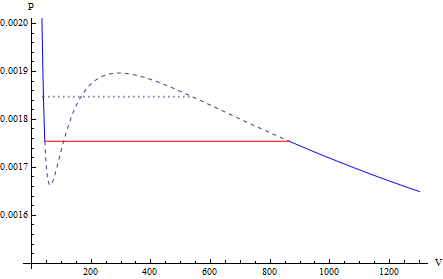}
\includegraphics[width=210pt]{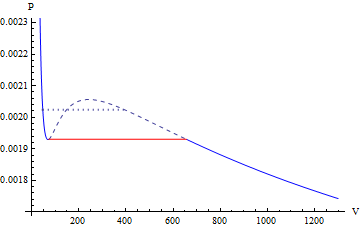}
\caption{{\it The plot of the isotherm for $T/T_c=0.9$ (left) and $T/T_c=0.93274820$ (right) in the $P-V$ diagram. The oscillating part of the isotherm should be replaced by an isobar (the red solid line), which is lower than the one given by the Maxwell's equal area law (the dotted line). Thus, the area above the isobar is larger than the one below it. If the temperature increases further, such that the isobar coincides with the tangent of the local minimum of the oscillating part (the right panel), it gives rise to the terminating point of the first order transition. We have set $q=1$.}}
\label{fig2}\end{figure}

\begin{figure}[ht]
\centering
\includegraphics[width=210pt]{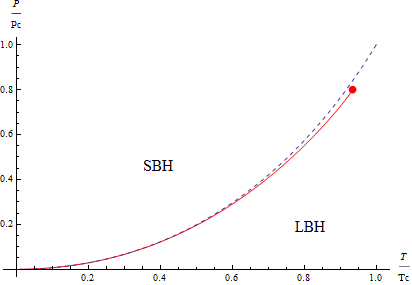}
\includegraphics[width=210pt]{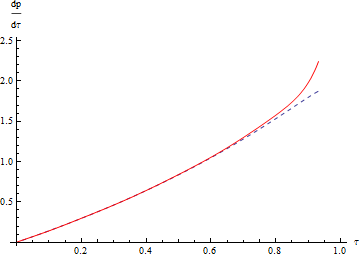}
\caption{{\it The left panel is the coexistence curve of the first order transition (solid line) which terminates at $T=T_*$ (red point). The curve is lower than the one (dashed line) given by Maxwell's equal area law. The right panel is the slope of the coexistence curve (solid line) during the transition. It is slightly higher than the one (dashed line) given by the Clapeyron equation.}}
\label{fig3}\end{figure}

In order to describe the SBH-LBH transition which is of first order and find its terminating point, let us first re-write the first law (\ref{firstlaw}) via a Legendre transformation
\be dG=-S dT+\Psi dQ_m+V dP+\Pi d\sigma \,.\label{firstlaw2}\ee
Recall that a first order transition occurs at a fixed temperature and pressure and both the coexistence phases have the same Gibbs free energy. Thus, in the $P-V$ plane, one has
\be 0=\Psi dQ_m+V dP+\Pi d\sigma  \,.\label{coe1}\ee
This is the physical condition that allows us to study the area law in the $P-V$ diagram\footnote{Of course, one can also write down the equation in the $T-S$ plane and parallelly discuss the corresponding area law.}. It is known that the oscillating part of the isotherm should be replaced by an isobar at the (first order) transition point (see Fig. \ref{fig2}). In general, if one works in a canonical ensemble with fixed $\sigma\,,Q_m$, then
the condition (\ref{coe1}) leads to the Maxwell's equal area law $\oint V \mathrm{d}P=0$: the areas above and below the isobar are equal one another. However, in our case when deriving the equation of state, we take $\sigma$ as a dynamical variable as well as the magnetic charge and the conjugate potential. Hence, the system is neither a canonical nor a grand canonical ensemble. Consequently, the Maxwell's equal area law is no loner valid in our case. The new area law turns out to be
\be \oint VdP=-\oint \Psi dQ_m-\oint \Pi d\sigma \,.\label{arealaw}\ee
The term on the r.h.s of this equation is non-vanishing and plays a central role in characterizing the details of the SBH-LBH transition. Solving this equation at a given temperature (below the terminating point), we can get the corresponding pressure and the volumes (or entropies) of the coexistent phases. We find that the pressure is smaller than that given by the Maxwell's equal area law (see Fig. \ref{fig2}). In other words, now the area above the isobar is larger than the one below it. The difference of the weight between the two areas increases with the temperature and arrives at a maximum when the isobar is tangential to the local minimum of the oscillating part. This exactly corresponds to the true critical point at which the first order transition terminates. We find
\be T_*/T_c= 0.93274820\,,\quad P_*/P_c=0.79608641\,,\quad \nu^*_S=0.83347099\,,\quad \nu^*_L=1.7693557 \,,\ee
where all the quantities associated with this point are denoted by a star. In Fig. \ref{fig3}, we plot the coexistence curve $p=p(\tau)$ of the first order transition. The curve was obtained by solving the area law (\ref{arealaw}) at a fixing temperature which runs over $0<T\leq T_*$. Alternatively, we can also find such a curve by requiring the two phases have the same Gibbs free energy with two different volumes. The two methods give the same result. It is clear that the curve is slightly lower (at most $5\%$) than that given by the Maxwell's equal area law. As pointed out in \cite{Wei:2014qwa}, the SBH-LBH coexistence curve has a parametric form
 \be p=\sum_{i>0}c_i \tau^i \,.\ee
For our case, we find the fitting formula
\bea p&=&0.785932 \tau^2-2.14636\tau^3+21.9592\tau^4-109.056\tau^5\nn\\
&+&315.680 \tau^6-550.668\tau^7+570.807\tau^8-323.835\tau^9+77.4554\tau^{10} \,,\label{fitting}\eea
perfectly matches with the numerical result. It should be emphasized that for vdW fluid, the coexistence curve was governed by the Clapeyron equation
 \be \fft{dP}{dT}=\fft{S_g-S_l}{V_g-V_l} \,,\label{clap}\ee
 while in our case this is no longer true.
 In Fig. \ref{fig3}, we also plot the slope of the coexistence curve in the right panel. The numerical result (solid) is slightly larger (at most $10\%$) than that given by the Clapyron equation. Hence, some temperature dependent corrections should be considered on the r.h.s of the Eq.(\ref{clap}). Unfortunately, we cannot derive these extra terms analytically from the first law since $\sigma$ is discontinuously changed during the transition. From Fig. \ref{fig3} we also observed that the Maxwell's equal area law as well as the Clapyron equation is approximately valid in the low temperature limit. A naive interpretation of this is that the system may approximate a canonical ensemble in the low temperature limit. However, this is not true. In fact, the change of the parameter $\sigma$ (or its inverse) and the magnetic charge in the transition increases
 as the temperature decreases. Nonetheless, its integrated effect, the weight of the terms on the r.h.s of (\ref{arealaw}) is sufficiently suppressed in the low temperature limit. The physical interpretation of this deserves further studies.

\begin{figure}[ht]
\centering
\includegraphics[width=210pt]{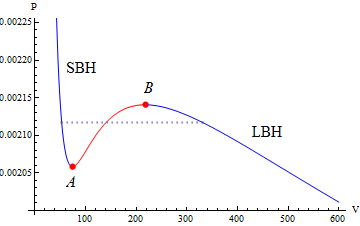}
\includegraphics[width=210pt]{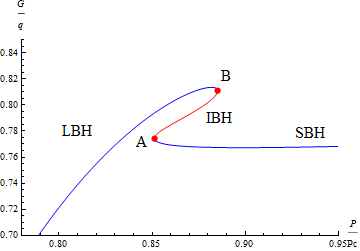}
\caption{{\it The left panel is an isotherm $T/T_c=0.95$ in the $P-V$ diagram. There still exists an oscillating part in the isotherm but one cannot find an isobar to saturate the area law (the dotted line is given by the Maxwell's equal area law). The right panel is the Gibbs free energy as a function of the pressure at the same temperature. The two turning points correspond to the local minimum and maximum of the isotherm in the $P-V$ diagram. The red line between the two turning points denotes some intermediate black holes (IBH). }}
\label{fig4}\end{figure}

When the temperature increases further such that $T_*<T<T_c$, one can no longer find an isobar to saturate the area law (\ref{arealaw}) in the $P-V$ diagram (see Fig. \ref{fig4}). Hence, the first order SBH-LBH transition can not happen any longer. In Fig. \ref{fig4}, we also plot the Gibbs free energy as a function of the pressure
for the same isotherm $T/T_c=0.95$. It is clear that when $T>T_*$, the characteristic ``swallow-tail" behavior of the Gibbs free energy for a first order transition disappears. Instead, an oscillating region emerges for a certain range of the pressure. The two turning points\footnote{The first and second order derivative of the Gibbs free energy with respect to the pressure diverge at the two turning points.} of the Gibbs free energy exactly correspond to the local minimum and maximum of the isotherm in the $P-V$ diagram, respectively. The three parts of the Gibbs free energy describing different phases of the black hole have their counterparts in the $P-V$ isotherms.
Therefore, the ``exotic" behavior of the Gibbs free energy is a direct consequence of the existence of oscillating isotherms in the $P-V$ diagram. This probably implies that when $T_*<T<T_c$, the system is in some stable mixture phases, although the small-large black hole phases can still be clearly distinguished from one another\footnote{ We have carefully checked that at least, a new transition of second order cannot happen in this case.}.


\subsection{Critical exponents}
For vdW fluid, there are various exponents characterizing the behavior of physical quantities near the critical point. They are defined by
\bea\label{expvdw}
&&C_v=T\fft{\partial S}{\partial T}\Big|_{v}\propto |t|^{-\alpha}\,,\nn\\
&&\eta=v_L-v_S\propto |t|^\beta \,,\nn\\
&&\kappa_T=-\fft{1}{V}\fft{\partial V}{\partial P}\Big |_T\propto |t|^{-\gamma} \,,\nn\\
&&|P-P_c|_{T=T_c}\propto |V-V_c|^\delta\,,
\eea
where $t=(T-T_c)/T_c$ (it should not be confused with the time coordinate); $C_v$ is the specific heat at constant volume; $\eta$ is the order parameter, which measures the differences of the specific volume between the gas (LBH) phase $v_L$ and the liquid (SBH) phase $v_S$; $\kappa_T$ is the isothermal compressibility. Note that in above definitions, $(T_c\,,P_c\,,V_c)$ denotes the true critical point at which the first order transition terminates with an unique thermodynamic volume. Thus, the definitions are valid for the transition satisfying the Maxwell's equal area law. On the contrary, in our case the equal area law is no longer valid and the first order phase transition terminates at the point $(T_*\,,P_*\,,V_S^*\,,V_L^*)$ which has discontinuous thermodynamic volumes. Therefore, the various exponents above should be redefined properly. We define
\bea\label{exponent}
&&C_v=T\fft{\partial S}{\partial T}\Big|_{v}\propto |t_*|^{-\alpha}\,,\nn\\
&&\eta=v_L-v_S \propto|t_*|^\beta \,,\nn\\
&&\kappa_T|_{\mathrm{SBH}}=-\fft{1}{V_S}\fft{\partial {V_S}}{\partial P}\Big |_T\propto |t_*|^{-\gamma_S} \,,\nn\\ &&\kappa_T|_{\mathrm{LBH}}=-\fft{1}{V_L}\fft{\partial {V_L}}{\partial P}\Big |_T\propto |t_*|^{-\gamma_L}\,,\nn\\
&&|P-P_*|_{T=T_*}\propto |V-V_S^*|^{\delta_S}\propto |V-V_L^*|^{\delta_L}\,,
\eea
where $t_*=(T-T_*)/T_*$. It should be emphasized that owing to the discontinuity of the thermodynamic volume at the critical point, we now in principle have two $\gamma$ and two $\delta$ exponents associated with the critical SBH phase and LBH phase respectively.

Let us compute the various exponents defined above (the computation of the exponents for AdS black hole systems satisfying the equal area law is briefly reviewed in the Appendix). First, the entropy is
\be S=\pi r_0^2=\fft{\pi}{4}\Big( \fft{6V}{\pi} \Big)^{2/3} \,,\ee
which is independent of $T$ at a constant volume. Thus $C_v=0$, implying that $\alpha=0$.
\begin{figure}[ht]
\centering
\includegraphics[width=230pt]{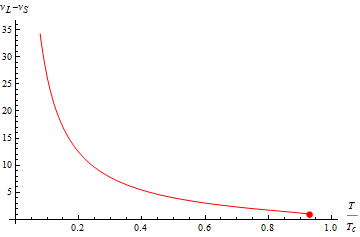}
\caption{{\it The order parameter $\eta$ is a smooth function of the temperature. The small red disk corresponds to the critical point $T=T_*$. }}
\label{fig5}\end{figure}
The order parameter $\eta$ is a smooth function of $T$ (see Fig. \ref{fig5}). Near the critical point, we have
\be  \eta=\eta_*+\eta_*' |t_*|+O(|t_*|^2)\qquad \Longrightarrow\qquad \beta=0 \,,\ee
where both $\eta_*=v_L^*-v_S^*$ and $\eta_*'=|\eta'(0)|$ take some finite value at the critical point. In order to calculate the exponents $\gamma$ and $\delta$, it is more convenient for us to define a new quantity $\omega=V/V_c-1=\nu^3-1$ and then rewrite the equation of state using $\omega$. Note that $\omega $ also has discrete values $(\omega_S^*\,,\omega_L^*)$ at the critical point. For simplicity, we collectively denotes the two critical values by $\omega_c$. Then for a small variation of the thermodynamic volume near the critical point, namely $\omega=\omega_c+\delta \omega$ ($\delta \omega<0$ on the SBH side and $\delta\omega>0$ on the LBH side), the law of the corresponding state can be expanded as follows:
\be p=p_*+a_{01}t_*+(a_{10}+a_{11}t_*)\delta\omega+(a_{20}+a_{21}t_*)\delta\omega^2+O(\delta\omega^3) \,,\label{corres}\ee
where $a_{ij}$ are constant coefficients, which in general have different values for the SBH and LBH phases respectively. It is worth pointing out that $a_{10}$ and $a_{20}$ vanish for vdW fluid while in our case
\bea
&& a_{10}|_{\mathrm{SBH}}=-1.59399\times 10^{-8}\,,\qquad a_{10}|_{\mathrm{LBH}}=-0.01738660 \,,\nn\\
&& a_{20}|_{\mathrm{SBH}}=0.51141800\,,\qquad a_{20}|_{\mathrm{LBH}}=0.00050091 \,.
\eea
 All of these are non-vanishing (One may worry about that the tiny value of $a_{10}|_{\mathrm{SBH}}$ comes from some numerical error. To exclude this possibility, we improve the working precision of the calculations and obtain numerical results with several higher orders of accuracy. We find that $a_{10}|_{\mathrm{SBH}}$ still has a finite value of the same order $10^{-8}$. In addition, for systems with vanishing $a_{10}$, our numerical result gives $a_{10}\sim 10^{-16}$. This gives us strong confidence that the above result is robust.). This is important in the derivation of $\gamma\,,\delta$ exponents. By definition, the isotherm compressibility can be calculated as follows
\bea \kappa_T&=&-\fft{1}{P_c(1+\omega)}\fft{\partial \omega}{\partial p}\Big |_{t_*}=-\fft{1}{P_c(1+\omega_f+\delta\omega)}\fft{\partial \delta\omega}{\partial p}\Big |_{t_*} \nn\\
&=&-\fft{1}{P_c(1+\omega_f+\delta\omega)(a_{10}+a_{11}t_*)}=-\fft{1}{P_c(1+\omega_f)a_{10}}+O(\delta\omega\,,t_*)\,.
\eea
Thus, we find
\be \gamma_S=\gamma_L=0 \,.\ee
Finally, at the critical isotherm with $t_*=0$, we have (to leading order)
\be p-p_*=a_{10}\delta \omega\qquad \Longrightarrow \qquad \delta_S=\delta_L=1 \,.\ee
In summary, we get
\be \alpha=0\,,\quad \beta=0\,,\quad \gamma_S=\gamma_L=0\,,\quad \delta_S=\delta_L=1\,.\label{exp} \ee
The various exponents are universal as well as the vdW fluid but now $\beta\,,\gamma\,,\delta$ have different values. This reflects the differences how the first order transition terminates at the critical point.

Finally, unlike the vdW fluid, the specific heat at constant pressure in our case is finite at the critical point. This can be directly seen from the
definition
\be C_p=T\fft{\partial S}{\partial T}\Big |_P \,,\ee
together with
\be T=\fft{1}{4\sqrt{\pi S}}\Big(1+\fft{8P S^{5/2}-3\pi^{3/2}q^3}{S^{3/2}+\pi^{3/2}q^3} \Big) \,.\ee
We find
\be C_p^*|_{SBH}=552.487 q^2\,,\qquad C_p^*|_{LBH}=576.755 q^2 \,.\ee
Nevertheless, a straightforward calculation shows that $C_p$ still diverges at the inflection point $T=T_c\,,P=P_c$.

\subsection{With singularities}

\begin{figure}[ht]
\includegraphics[width=210pt]{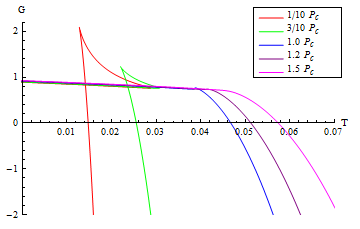}
\includegraphics[width=210pt]{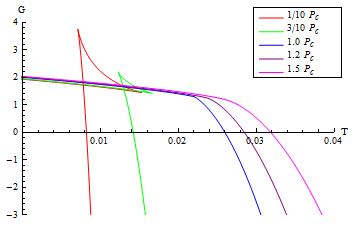}
\caption{{\it The Gibbs free energy is plotted as a function of the temperature for $M=1/10$ (left) and $M=-10$ (right) respectively. We have set $q=1$.}}
\label{figs1}\end{figure}
\begin{figure}[ht]
\includegraphics[width=210pt]{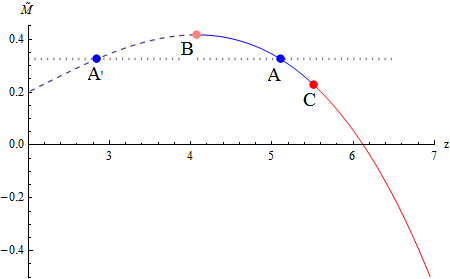}
\includegraphics[width=210pt]{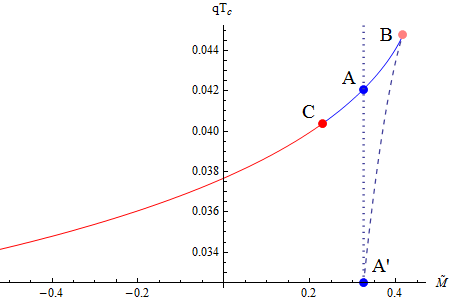}
\caption{{\it The plot of the Schwarzschild mass $\widetilde{M}$ as a function of $z$ and the critical temperature $T_c$ as a function of $\widetilde{M}$. The solid lines correspond to the physical parameter space.}}
\label{figs2}\end{figure}

For the more general solution (\ref{sol2}), the equation of state receives an extra term, which is proportional to the {\it Schwarzschild mass}
\be P=\fft{T}{v}-\fft{1}{2\pi v^2}+\fft{8T q^3}{v^4}+\fft{8q^3}{\pi v^5}-\fft{48 M q^3}{\pi v^6} \,.\label{geneeos}\ee
Here we also take $\sigma$ as a dynamical variable. Of course, one can also study the criticality for a canonical ensemble (with fixed $\sigma$ and $Q_m$) since the solution has one more free parameter. This will be discussed in the next section. By plotting the $P-V$ diagram, we find that an oscillating part exists for the isotherm below some critical temperature. Moreover,
from the characteristic behavior of the Gibbs free energy (see Fig. \ref{figs1}), we easily confirm that a first order transition occurs when the temperature $T<T_*<T_c$ for both positive and negative {\it Schwarzschild mass}. Here the critical points $T_*\,,T_c$ are defined as before. By definition (\ref{crit}), we first obtain
\be P_c=\fft{5z^6-112z^3-64}{3\pi q^2 z^5(5z^3+64)}\,,\quad T_c=\fft{4(z^3-10)}{\pi q z(5z^3+64)}\,,\quad v_c=q z \,,\ee
and
\be \fft{P_c v_c}{T_c}=\fft{5z^6-112z^3-64}{12z^3(z^3-10)} \,,\ee
where $z$ is a function of the dimensionless ratio $\widetilde{M}=M/q$, satisfying an algebraic equation
\be z(z^6-224z^3-1280)+288\widetilde{M}(5z^3+64)=0  \,.\ee
In Fig. \ref{figs2}, we plot $\widetilde{M}$ as a function of $z$. It is clear that in general, $z$ is not a single valued function of $\widetilde{M}$. However, we shall require the critical temperature $T_c$ and pressure $P_c$ being real and positive, which gives $z>z_{A'}=2.84212\,,\widetilde{M}\leq \widetilde{M}_B=0.417041$. This is the least condition that a small-large black hole transition occurs. In addition, for any given value in $\widetilde{M}_A<\widetilde{M}< \widetilde{M}_B$, there exists two valid $z$ and hence two critical temperature $T_c$. However, the lower one in fact is nonphysical. Therefore, we conclude that the physical parameter space is $z\geq z_B$, in which $\widetilde{M}$ is a monotonous decreasing function of $z$.

Furthermore, we find that there exists an upper bound $\widetilde{M}=\widetilde{M}_C\approx 0.22\sim 0.23$ at which the $P-V$ criticality disappears.
This is a direct consequence of the fact that the {\it Schwarzschild mass} term in the equation of state (\ref{geneeos}) acquires a negative sign. For any positive $M$, there exists a competition between this term and the third and forth terms. When the latter terms are dominant, the $P-V$ criticality is governed while when the $M$ term dominates, the criticality is ruined. Such a competing phenomena was also observed in Gauss-Bonnet gravity \cite{Cai:2013qga}.
\begin{table}[h]
\centering
\begin{tabular}{|c|c|c|c|c|c|c|c|c|}
  \hline
  $\widetilde{M}$ & 0.22 & 0 & -1.5 & -4.5  \\ \hline
  $T_*/T_c$ & 0.93045990 & 0.93274820 & 0.94618735 & 0.95765300 \\ \hline
  $P_*/P_c$ & 0.77595051 & 0.79608641 & 0.85072948 & 0.88662577  \\ \hline
  $\widetilde{M}$ & -18 & -100 &-500 & -1000000 \\ \hline
  $T_*/T_c$ &  0.97273625 & 0.98624210 & 0.99327870 & 0.99982994 \\ \hline
  $P_*/P_c$ &  0.92937801 & 0.96528246 & 0.98330652 & 0.99958962   \\ \hline
\end{tabular}
\caption{The critical temperature and pressure $(T_*\,,P_*)$ increases as the {\it Schwarzschild mass} decreases. They approach the critical point $(T_c\,,P_c)$ when $\widetilde{M}$ becomes sufficiently negative.}
\label{critstar}
\end{table}
In contrast, for negative $M$ there does not exist any competition between these terms. However, to avoid ghost-like graviton modes, we shall require the enthalpy (or the AMD mass) being positive definite, which may lead to some lower bound for $\widetilde{M}$. We find
\be H=\mathrm{Positive\,\, Terms}-\fft{8\widetilde{M}}{z^3\nu^3} \,,\ee
where ``Positive Terms" denotes the terms that are always positive in the parameter space. It is clear that for negative $\widetilde{M}$, the enthalpy is always positive definite. Thus we safely conclude that the {\it Schwarzschild mass} does not have a lower bound.

In the physical parameter space with $P-V$ criticality, we observe that (from Table \ref{critstar}) the terminating point $(T_*\,,P_*)$ of the transition increases as the {\it Schwarzschild mass} $\widetilde{M}$ decreases and the ratios $T_*/T_c\,,P_*/P_c$ approach unity in the large $|\widetilde{M}|$ limit.

Finally, we find that the various exponents defined by (\ref{exponent}) turn out to be universal. They are still given by (\ref{exp}), independent of the {\it Schwarzschild mass}. The computation of these exponents follows the method demonstrated in the last subsection. The only difference is that the various coefficients $a_{ij}$ of the expansion (\ref{corres}) of the corresponding state are now functions of $z$. In particular, we carefully check that $a_{10}(z)$ is always nonzero in both SBH and LBH phases.

\section{P-V criticality for canonical ensemble}
As mentioned earlier, for the general black hole solution (\ref{sol2}) we can also study the $P-V$ criticality for fixed $\sigma$ and magnetic charge $Q_m$. The equation of state becomes
\be\label{eosce}
P=\fft{T}{v}-\fft{1}{2\pi v^2}+\fft{48\sigma^{-1}q^6}{\pi\big( v^3+8q^3 \big)^{2}}\,.
\ee
\begin{figure}[ht]
\includegraphics[width=210pt]{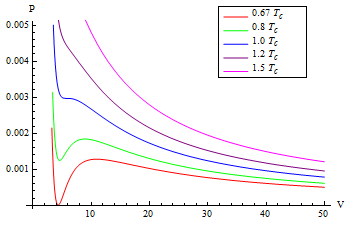}
\includegraphics[width=210pt]{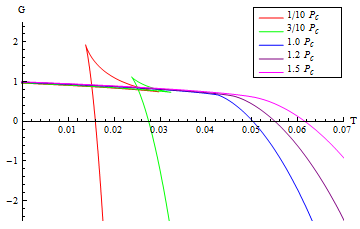}
\caption{{\it The P-V diagram (left) and Gibbs free energy (right) for the generalized Hayward black hole with fixed $\sigma$ and $q$. We have set $\widetilde{\sigma}=1\,,q=1$. }}
\label{figc1}\end{figure}
The isotherms of the $P-V$ diagram and the Gibbs free energy as a function of the temperature are depicted in Fig. \ref{figc1}. The characteristic behaviors in both plots indicate that a SBH-LBH transition of first order can occur below some critical temperature.

\begin{figure}[ht]
\includegraphics[width=210pt]{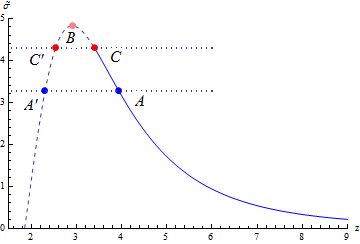}
\includegraphics[width=210pt]{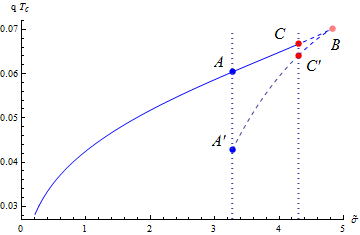}
\caption{{\it The left panel is $\widetilde{\sigma}$ as a function of $z$ whilst the right panel is the critical temperature as a function of $\widetilde{\sigma}$. The solid lines correspond to the physical parameter space with critical phenomena.}}
\label{figc2}\end{figure}

Notice that the Maxwell's equal area law now is saturated because we work in a canonical ensemble. Thus, the critical point exactly coincides with the inflection point of the isotherms. We get
\be P_c=\fft{5z^6-64z^3+32}{3\pi q^2 z^5(5z^3-32)}\,,\qquad T_c=\fft{4(z^3-10)}{\pi q z(5z^3-32)}\,,\qquad \fft{P_c v_c}{T_c}=\fft{5z^6-64z^3+32}{12z^3(z^3-10)} \,,\ee
where $z=v_c/q$ is a function of the dimensionless ratio $\widetilde{\sigma}=\sigma/q^2$, given by
\be (z^3+8)^4-288\widetilde{\sigma}^{-1}z^5(5z^3-32)=0 \,.\ee
In general, $z$ and $T_c$ are not single valued functions of $\widetilde{\sigma}$ (see Fig. \ref{figc2}). However, a detailed analysis shows that in the parameter space with $P-V$ criticality, they are monotone functions of $\widetilde{\sigma}$. First, the critical temperature and pressure should be real and positive which gives $z>z_{A'}=2.30702\,,\widetilde{\sigma}\leq \widetilde{\sigma}_B= 4.82824$. This is a least condition. Furthermore, $\widetilde{\sigma}$ has an upper bound around $4.2\sim 4.3$ (corresponding to $C\,,C'$) at which the criticality disappears. This is easily understood because in the large $\sigma$ limit, the general black hole solution (\ref{sol2}) becomes a Schwarzschild black hole, which does not allow any critical phenomena. This can also be seen directly from the equation of state (\ref{eosce}).
In addition, for $\widetilde{\sigma}_A<\widetilde{\sigma}<\widetilde{\sigma}_C$, the critical temperature is apparently double valued. However, the lower value turns out to be nonphysical. Thus, to govern the $P-V$ criticality, $z$ should take values as $z>z_C$ and the critical point is uniquely defined.

\begin{figure}[ht]
\includegraphics[width=210pt]{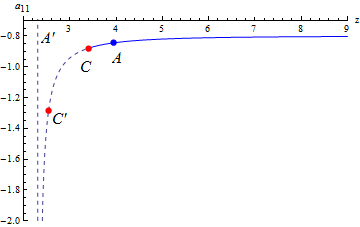}
\includegraphics[width=210pt]{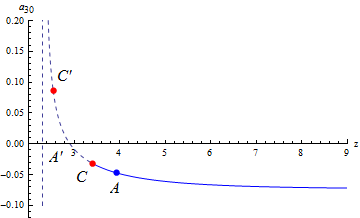}
\caption{{\it The plots for $a_{11}$ (left) and $a_{30}$ (right) as a function of $z$.}}
\label{figc3}\end{figure}

A standard calculation (see the Appendix) shows that the various exponents defined by (\ref{expvdw}) coincide with those of vdW fluid, namely
\be \alpha=0\,,\quad \beta=1/2\,,\quad \gamma=1\,,\quad \delta=3 \,.\ee
In particular, as the specific values of $\beta\,,\gamma\,,\delta$ strongly depend on the behavior of the  corresponding state near the critical point, let us discuss this further. We find
\be p=1+a_{01}\, t+a_{11}\, t\omega+a_{30}\, \omega^3+O(t\omega^2\,,\omega^4) \label{pexpand1}\,,\ee
where the coefficients $a_{ij}$ are all functions of $z$, for example
\bea &&a_{01}=\fft{12z^3(z^3-10)}{5z^6-64z^3+32}\,,\qquad a_{11}=-\fft{4z^3(z^3-10)}{5z^6-64z^3+32}\,,\nn\\
&& a_{30}=-\fft{2z^3(5z^6-136z^3+320)}{27(z^3+8)(5z^6-64z^3+32)} \,.\eea
It is worth pointing out that now $a_{10}\,,a_{20}$ strictly vanish such that $\gamma\,,\delta$ will have different values from (\ref{exp}). In the parameter space with $P-V$ criticality $z>z_C$, $a_{10}(z)$ is
always positive whilst $a_{11}(z)$ and $a_{30}(z)$ are always negative (see Fig. \ref{figc3}). Thus the calculation of $\beta$ which requires $a_{11}/a_{30}>0$ is valid in this case.
The specific value $\beta=1/2<1$ in fact reflects that the derivative of the order parameter $\eta$ at the critical point diverges as $\eta'\propto |t|^{-1/2}$.

\section{Conclusion}
In this paper, we consider Einstein gravity with a negative cosmological constant coupled to a certain non-linear electrodynamics. We obtain the magnetically charged static spherical symmetric black hole solution with two independent free parameters. In the neutral limit, the solution reduces to a Schwarzschild black hole. In particular, for a degenerate case with zero {\it Schwarzschild mass} the solution is the Hayward black hole generalized in AdS space-time. We study the curvature polynomials and find that in this case the geometry is regular everywhere in the space-time.

We then study the global properties of the solution and derive the first law of thermodynamics. Treating the cosmological constant and the parameter $\sigma$ associated with the non-linear electrodynamics as thermodynamical variables, we further generalize the first law in the extended phase space and obtain the corresponding Smarr formula, which is consistent with scaling dimensional argument.

 Furthermore, we study the $P-V$ criticality of Hayward-AdS black hole in the extended phase space by taking $\sigma$ as a dynamical variable. From the characteristic behaviors of the isotherms in the $P-V$ diagram and the Gibbs free energy, we observe that there exists a first order small-large black hole transition below some critical temperature. However, the system has many intriguing properties that differ from the vdW fluid. First, the Maxwell's equal area law in the $P-V$ (or $S-T$) diagram is no longer valid. The new area law gives an isobar that is lower than the one given by the equal area law in the isotherms. As a consequence, the critical point of the first order transition ($T_*\,,P_*$) does not coincide with the inflection point ($T_c\,,P_c$) of the isotherms. We find $T_*/T_c<1\,,P_*/P_c<1$. This leads to many interesting results such as: the coexistence curve of the transition can not be characterized by the Clapyron equation; the various exponents characterizing the behaviors of the physical quantities near the critical point are different from those of vdW fluid and the specific heat at constant pressure is finite at the critical point. In addition, there exits a region $T_*<T<T_c$ at which the smarr-large black hole transition cannot be first order. It is an open problem whether there exists a new transition between small-large black hole phases in this case.

We then study the critical phenomena for the general black hole solution with nonzero {\it Schwarzschild mass} $M$. In this case, the equation of state receives an extra term proportional to $-M$. The fact that this term acquires a negative sign leads to a competition between this term and the other terms in the equation of state. Consequently, for positive $M$ these exists an upper bound at which the criticality disappears. Moreover, we find that the critical temperature and pressure ($T_*\,,P_*$) increases as the {\it Schwarzschild mass} decreases and approach the inflection point ($T_c\,,P_c$) in the large $|M|$ limit.

In the end, we also briefly discuss the critical phenomena of the general black hole solution with fixed $\sigma\,,Q_m$. In this case, the system is a canonical ensemble and the properties of the dual fluid roughly coincide with those of the vdW fluid. 

Recently, it was established \cite{Johnson:2013dka,Caceres:2015vsa,Nguyen:2015wfa,Zeng:2015tfj,Zeng:2015wtt,Zeng:2016sei,Mo:2016cmi} that some non-local observables can successfully capture the information of the extended phase structure. We leave the analysis of these as a possible future direction for research.

\section*{Acknowledgments}
 This work was in part supported by NSFC Grants No.~11275010, No.~11335012 and No.~11325522.

\section{Appendix: Critical exponents of of AdS black hole systems satisfying Maxwell's equal area law}

For vdW fluid, there are various critical exponents characterizing the behavior of physical quantities near the critical points. They are defined by (\ref{expvdw})
\bea
&&C_v=T\fft{\partial S}{\partial T}\Big|_{v}\propto |t|^{-\alpha}\,,\nn\\
&&\eta=v_l-v_s\propto |t|^\beta \,,\nn\\
&&\kappa_T=-\fft{1}{V}\fft{\partial V}{\partial P}\Big |_T\propto |t|^{-\gamma} \,,\nn\\
&&|P-P_c|\propto |V-V_c|^\delta\,,\nn
\eea
where $t=(T-T_c)/T_c$. Following the method established in \cite{Kubiznak:2012wp}, one can compute the various exponents for the AdS black hole systems satisfying the Maxwell's equal area law straightforwardly. Here we focus on the charged black holes in Einstein (or Gauss-Bonnet) gravity minimally coupled to matter fields.

First, the entropy is only a function of the thermodynamic volume. For example, in four dimensions
\be S=\pi r_0^2=\fft{\pi}{4}\Big( \fft{6V}{\pi} \Big)^{2/3} \,,\ee
is independent of $T$ at constant volume. Hence $C_v=0$, implying that $\alpha=0$.

The computation of the other exponents $\beta\,,\gamma$ and $\delta$ strongly depends on the series expansion of the equation of state near the critical point. Defining
\be p=P/P_c\,,\qquad \nu=v/v_c\,,\qquad \omega=V/V_c-1=\nu^3-1 \,,\ee
one finds
\be p=1+a_{10}\, t+a_{11}\, t\omega+a_{30}\, \omega^3+O(t\omega^2\,,\omega^4) \label{pexpand}\,,\ee
where $a_{ij}$ are constant coefficients. As an example, for Reissner-Nordstr\"{o}m black holes one has
\be a_{10}=8/3\,,\qquad a_{11}=-8/9\,,\qquad a_{30}=-4/81 \,.\ee
Note that the above expansion has been cut off at relevant orders. The validity of this could be justified by Eq.(\ref{wls}) below. Differentiating the expansion for fixed $t<0$, we obtain
\be dP=P_c(a_{11}\, t+3a_{30}\,\omega^2) \,.\ee
Then using the Maxwell's equal area law $\oint V dP=0$, we get two equations
\bea
&&p=1+a_{10}\,t+a_{11}\, t\omega_s+a_{30}\,\omega_s^3=1+a_{10}\,t+a_{11}\, t\omega_l+a_{30}\,\omega_l^3\,,\nn\\
&&0=\int_{\omega_s}^{\omega_l} \omega(a_{11}\, t+3a_{30}\,\omega^2)\,.
\eea
The unique non-trivial solution is
\be \omega_l=-\omega_s=\sqrt{-\fft{a_{11}\,t}{a_{30}}} \,.\label{wls}\ee
Note that since $t<0$, the reality of $\omega_{l,s}$ requires $a_{11}/a_{30}$ being positive (this is easily verified for RN black holes). We obtain
\be \eta=V_c(\omega_l-\omega_s)=2V_c\sqrt{-\fft{a_{11} t}{a_{30}}}\propto |t|^{1/2} \qquad \Longrightarrow\qquad \beta=1/2 \,.\ee
To calculate $\gamma$, we have
\be \kappa_T=-\fft{1}{V}\fft{\partial V}{\partial P}\Big|_T=-\fft{1}{P_c(1+\omega)}\fft{\partial \omega}{\partial p}\Big|_t=-\fft{1}{a_2 P_c t}+O(\omega)
\qquad \Longrightarrow\qquad \gamma=1 \,.\ee
Finally, the definition of $\delta $ is equivalent to $|p-1|\propto \omega^\delta$ when $t=0$. Hence, from the expansion (\ref{pexpand}) we find
$\delta=3$. In summary, the various exponents are
\be \alpha=0\,,\quad \beta=1/2\,,\quad \gamma=1\,,\quad \delta=3 \,,\label{exponents}\ee
which perfectly coincide with those of the vdW fluid.
It is worth pointing out that the calculation of the exponents $\beta\,,\gamma\,,\delta$ strongly depends on the expansion of the corresponding state near the critical point. Once the expansion (\ref{pexpand}) is given, one will arrive at above results for the three exponents. This is independent of the precise value of the series coefficient $a_{ij}$ but one should remind that $a_{11}/a_{30}$ should be positive definite.


\begin{thebibliography}{100}


\bibitem{hawking}
  S.W.~Hawking and G.F.R.~Ellis,
{\it The Large Scale Structure of Spacetime} (Cambridge University Press, Cambridge 1973).

\bibitem{bardeen}
J.M.~Bardeen, in: Conference Proceedings of GR5, Tbilisi, USSR, 1968, p. 174.


\bibitem{Borde:1994ai}
  A.~Borde,
  {\it Open and closed universes, initial singularities and inflation,}
  Phys.\ Rev.\ D {\bf 50}, 3692 (1994).


\bibitem{Barrabes:1995nk}
  C.~Barrabes and V.~P.~Frolov,
  {\it How many new worlds are inside a black hole?,}
  Phys.\ Rev.\ D {\bf 53}, 3215 (1996).


\bibitem{Cabo:1997rm}
  A.~Cabo and E.~Ayon-Beato,
  {\it About black holes without trapping interior,}
  Int.\ J.\ Mod.\ Phys.\ A {\bf 14}, 2013 (1999).


\bibitem{Hayward:2005gi}
  S.~A.~Hayward,
  {\it Formation and evaporation of regular black holes,}
  Phys.\ Rev.\ Lett.\  {\bf 96}, 031103 (2006).


\bibitem{Bambi:2013ufa}
  C.~Bambi and L.~Modesto,
  {\it Rotating regular black holes,}
  Phys.\ Lett.\ B {\bf 721}, 329 (2013).


\bibitem{Ghosh:2014hea}
  S.~G.~Ghosh and S.~D.~Maharaj,
  {\it Radiating Kerr-like regular black hole,}
  Eur.\ Phys.\ J.\ C {\bf 75}, 7 (2015).

\bibitem{Toshmatov:2014nya}
  B.~Toshmatov, B.~Ahmedov, A.~Abdujabbarov and Z.~Stuchlik,
  {\it Rotating Regular Black Hole Solution,}
  Phys.\ Rev.\ D {\bf 89}, no. 10, 104017 (2014).


\bibitem{Azreg-Ainou:2014pra}
  M.~Azreg-A\"{\i}nou,
  {\it Generating rotating regular black hole solutions without complexification,}
  Phys.\ Rev.\ D {\bf 90}, no. 6, 064041 (2014).


\bibitem{AyonBeato:1998ub}
  E.~Ayon-Beato and A.~Garcia,
  {\it Regular black hole in general relativity coupled to nonlinear electrodynamics,}
  Phys.\ Rev.\ Lett.\  {\bf 80}, 5056 (1998).


\bibitem{AyonBeato:1999ec}
  E.~Ayon-Beato and A.~Garcia,
  {\it Nonsingular charged black hole solution for nonlinear source,}
  Gen.\ Rel.\ Grav.\  {\bf 31}, 629 (1999).


\bibitem{AyonBeato:1999rg}
  E.~Ayon-Beato and A.~Garcia,
  {\it New regular black hole solution from nonlinear electrodynamics,}
  Phys.\ Lett.\ B {\bf 464}, 25 (1999).

\bibitem{AyonBeato:2000zs}
  E.~Ayon-Beato and A.~Garcia,
  {\it The Bardeen model as a nonlinear magnetic monopole,}
  Phys.\ Lett.\ B {\bf 493}, 149 (2000).


\bibitem{Junior:2015fya}
  E.~L.~B.~Junior, M.~E.~Rodrigues and M.~J.~S.~Houndjo,
  {\it Regular black holes in $f(T)$ Gravity through a nonlinear electrodynamics source,}
  JCAP {\bf 1510}, 060 (2015).


\bibitem{fangw}
Zhong-Ying Fan, Sijie Gao and Xiaobao Wang,
{\it Regular Black Holes in General Relativity,}
in preparation.



\bibitem{Ashtekar:1984zz}
  A.~Ashtekar and A.~Magnon,
  {\it Asymptotically anti-de Sitter space-times,}
  Class.\ Quant.\ Grav.\  {\bf 1}, L39 (1984).


\bibitem{Ashtekar:1999jx}
A.~Ashtekar and S.~Das, {\it Asymptotically anti-de Sitter
space-times: Conserved quantities,} Class.\ Quant.\ Grav.\  {\bf
17}, L17 (2000) [arXiv:hep-th/9911230].


\bibitem{Rasheed:1997ns}
  D.~A.~Rasheed,
  {\it Nonlinear electrodynamics: Zeroth and first laws of black hole mechanics,}
  hep-th/9702087.


\bibitem{Kastor:2009wy}
  D.~Kastor, S.~Ray and J.~Traschen,
  {\it Enthalpy and the Mechanics of AdS Black Holes,}
  Class.\ Quant.\ Grav.\  {\bf 26}, 195011 (2009).


\bibitem{Cvetic:2010jb}
  M.~Cvetic, G.~W.~Gibbons, D.~Kubiznak and C.~N.~Pope,
  {\it Black Hole Enthalpy and an Entropy Inequality for the Thermodynamic Volume,}
  Phys.\ Rev.\ D {\bf 84}, 024037 (2011).


\bibitem{Kubiznak:2012wp}
  D.~Kubiznak and R.~B.~Mann,
  {\it P-V criticality of charged AdS black holes,}
  JHEP {\bf 1207}, 033 (2012).


\bibitem{Gunasekaran:2012dq}
  S.~Gunasekaran, R.~B.~Mann and D.~Kubiznak,
  {\it Extended phase space thermodynamics for charged and rotating black holes and Born-Infeld vacuum polarization,}
  JHEP {\bf 1211}, 110 (2012).


\bibitem{Banerjee:2011cz}
  R.~Banerjee and D.~Roychowdhury,
  {\it Critical phenomena in Born-Infeld AdS black holes,}
  Phys.\ Rev.\ D {\bf 85}, 044040 (2012).


\bibitem{Banerjee:2012zm}
  R.~Banerjee and D.~Roychowdhury,
  {\it Critical behavior of Born Infeld AdS black holes in higher dimensions,}
  Phys.\ Rev.\ D {\bf 85}, 104043 (2012).



\bibitem{Mo:2014qsa}
  J.~X.~Mo and W.~B.~Liu,
  {\it $P-V$ criticality of topological black holes in Lovelock-Born-Infeld gravity,}
  Eur.\ Phys.\ J.\ C {\bf 74}, no. 4, 2836 (2014).


\bibitem{Mo:2016jqd}
  J.~X.~Mo, G.~Q.~Li and X.~B.~Xu,
  {\it Effects of power-law Maxwell field on the critical phenomena of higher dimensional dilaton black holes,}
  Phys.\ Rev.\ D {\bf 93}, no. 8, 084041 (2016).

\bibitem{Hendi:2014kha}
  S.~H.~Hendi, S.~Panahiyan and B.~Eslam Panah,
  {\it P-V criticality and geometrical thermodynamics of black holes with Born-Infeld type nonlinear electrodynamics,}
  Int.\ J.\ Mod.\ Phys.\ D {\bf 25}, no. 01, 1650010 (2015).


\bibitem{Wei:2014qwa}
  S.~W.~Wei and Y.~X.~Liu,
  {\it Clapeyron equations and fitting formula of the coexistence curve in the extended phase space of charged AdS black holes,}
  Phys.\ Rev.\ D {\bf 91}, no. 4, 044018 (2015).


\bibitem{Cai:2013qga}
  R.~G.~Cai, L.~M.~Cao, L.~Li and R.~Q.~Yang,
  {\it P-V criticality in the extended phase space of Gauss-Bonnet black holes in AdS space,}
  JHEP {\bf 1309}, 005 (2013).
  
  
\bibitem{Johnson:2013dka}
  C.~V.~Johnson,
  {\it Large N Phase Transitions, Finite Volume, and Entanglement Entropy,}
  JHEP {\bf 1403}, 047 (2014).
  
  
\bibitem{Caceres:2015vsa}
  E.~Caceres, P.~H.~Nguyen and J.~F.~Pedraza,
  {\it Holographic entanglement entropy and the extended phase structure of STU black holes,}
  JHEP {\bf 1509}, 184 (2015).
  

\bibitem{Nguyen:2015wfa}
  P.~H.~Nguyen,
  {\it An equal area law for holographic entanglement entropy of the AdS-RN black hole,}
  JHEP {\bf 1512}, 139 (2015).

\bibitem{Zeng:2015tfj}
  X.~X.~Zeng, H.~Zhang and L.~F.~Li,
  {\it Phase transition of holographic entanglement entropy in massive gravity,}
  Phys.\ Lett.\ B {\bf 756}, 170 (2016).
  
\bibitem{Zeng:2015wtt}
  X.~X.~Zeng and L.~F.~Li,
  {\it Van der Waals phase transition in the framework of holography,}
  arXiv:1512.08855 [hep-th].
  
\bibitem{Zeng:2016sei}
  X.~X.~Zeng, X.~M.~Liu and L.~F.~Li,
  {\it Phase structure of the Born-Infeld-anti-de Sitter black holes probed by non-local observables,}
  arXiv:1601.01160 [hep-th].
  
  
\bibitem{Mo:2016cmi}
  J.~X.~Mo, G.~Q.~Li, Z.~T.~Lin and X.~X.~Zeng,
  {\it Van der Waals like behavior and equal area law of two point correlation function of f(R) AdS black holes,}
  arXiv:1604.08332 [gr-qc].
  
\end{thebibliography}
\end{document}